\documentclass{mn}

\usepackage{amsfonts}
\usepackage{amsmath}
\usepackage{graphicx}
 \font\sevenrm=cmr7 scaled 1000
\def\gsim{\;\lower4pt\hbox{${\buildrel\displaystyle >\over\sim}$}\;}
\def\lsim{\;\lower4pt\hbox{${\buildrel\displaystyle <\over\sim}$}\;}
\def\grls{\;\lower4pt\hbox{${\buildrel\displaystyle >\over <}$}\;}
\begin{document}
\title[Multi-Wavelength Correlations]
{Correlations among Multi-Wavelength 
Luminosities of Star-Forming Galaxies}
\author[Y.-Q. Lou and F.-Y. Bian]
       {Yu-Qing Lou$^{1,2,3}$ and
        Fu-Yan Bian$^{1}$\\
        $^1$Physics Department and the Tsinghua Center for
Astrophysics (THCA), Tsinghua University, Beijing 100084, China\\
        $^2$National Astronomical Observatories, Chinese Academy
of Sciences, A20, Datun Road, Beijing, 100012 China\\
        $^3$Department of Astronomy and Astrophysics, The
University of Chicago, 5640 South Ellis Ave., Chicago, 
IL 60637 USA. }
\date{Accepted 2004 ............;
      Received 2003.............;
      in original form 2003......}
%
                               
\pagerange{\pageref{firstpage}--\pageref{lastpage}} \pubyear{2004}

\maketitle

\label{firstpage}

\begin{abstract}
It has been known for two decades that a tight correlation 
exists between global far-infrared (FIR) and radio continuum
(1.4 and 4.8 GHz) fluxes/luminosities from galaxies,
which may be explained by formation activities of massive stars 
in these galaxies. For this very reason, a correlation might 
also exist between X-ray and FIR/radio global luminosities of 
galaxies. We analyze data from {\it ROSAT} All-Sky Survey and 
{\it IRAS} to show that such correlation does indeed exist between 
FIR ($42.5-122.5\mu$m) and soft X-ray ($0.1-2.4$keV) luminosities 
$L_{\rm X}$ and $L_{\rm FIR}$ in 17 normal star-forming galaxies 
(NSFGs) including 16 late-type galaxies and 1 host-dominant Seyfert 
galaxy as well as in 14 active star-forming galaxies (ASFGs) 
consisting of starburst-dominant Seyfert galaxies. The quantitative 
difference in such correlations in NSFGs and in ASFGs may be 
interpreted in terms of evolutionary variations from classic 
starburst galaxies to starburst-dominant Seyfert galaxies. 
Meanwhile, some low-luminosity AGNs (LLAGNs) tend to exhibit 
such a correlation that we infer for star-forming galaxies, 
implying that star-forming activities might be more dominant 
in LLAGNs. In contrast, AGN-dominant Seyfert galaxies do not 
show such a $L_{\rm X}$ versus $L_{\rm FIR}$ correlation;
this is most likely related to accretions towards 
supermassive black holes (SMBHs) in galactic nuclei. 
In order to establish a physical 
connection between the $L_{\rm X}-L_{\rm FIR}$ correlation and 
global star formation rate (SFR) in galaxies, we empirically derive 
both $L_{\rm X}-L_{\rm B}$ and $L_{\rm FIR}-L_{\rm B}$ relations 
with the blue-band luminosity $L_{\rm B}$ roughly representing the 
mass of a galaxy. It appears that the more massive galaxies are, the 
more star formation regions exist in these galaxies. The global SFR 
is not only associated with the mass of a galaxy but also closely 
related to the level of star-forming activities therein. We propose 
a relation between soft X-ray luminosity and SFR in star-forming 
galaxies. In order to probe the $L_{\rm X}-L_{\rm FIR}$ relation, 
we construct an empirical model in which both FIR and X-ray emissions 
consist of two components with one being closely associated with star 
formation and the other one not. Based on this model, we infer a 
linear relation between FIR/soft X-ray in star formation regions and 
radio luminosities, and get a linear relation between $L_{\rm X}$ and 
$L_{\rm FIR}$ for star forming regions. 
\end{abstract}

\section{Introduction}

A strong correlation exists between far-infrared (FIR) and radio
continuum (at 1.4 and 4.8 GHz frequencies) global fluxes/luminosities 
in star-forming galaxies as revealed by {\it Infrared Astronomy 
Satellite (IRAS)} data and radio continuum observations (e.g., 
Dickey \& Salpeter 1984; de Jong, Klein, Wielebinski \& Wunderlich 
1985; Helou, Soifer \& Rowan-Robinson 1985; Hummel et al. 1988;
Andersen \& Owen 1995; Condon, Anderson \& Helou 1991). At the 
time of this important discovery, this tight 
global correlation was unexpected because the radio and FIR emissions 
were thought to involve distinctly different physical processes, that 
is, the radio continuum emission is primarily caused by synchrotron 
emissions from relativistic cosmic-ray electrons gyrating around 
galactic magnetic fields (e.g., Lou \& Fan 2003 and references 
therein), while the FIR emission is primarily caused by thermal 
emissions from dust grains submerged in an intense ultraviolet (UV) 
radiation field (e.g., Lou \& Fan 2000 and references therein). 
Currently, the qualitative physical explanation for such a global 
correlation is that the FIR emission is mainly caused by dust grain 
absorption of UV photons emitted by young massive stars nearby, while 
sources of relativistic cosmic-ray electrons are mainly associated 
with magnetohydrodynamic (MHD) shocks of massive stellar winds or
supernova explosions (e.g., Lou 1993, 1994) that are end stages of 
massive stars --- the primary source of UV radiations that heat up 
interstellar dust grains. Globally, both FIR and radio emissions are 
thought to be intimately related to formation activities of massive 
stars in star-forming galaxies. This perspective is also supported 
by the fact that radio continuum emissions are well correlated with 
H$_{\alpha}$ emissions mainly from H{\sevenrm II} regions ionized 
by young massive stars (e.g., Lequeux 1971; Rieke \& Lebofsky 1979; 
Klein 1982; Wynn-Williams 1982; Hummel et al. 1988; Anderson \& 
Owen 1995; Lou \& Fan 2000, 2003; Lou et al. 2002; Lou \& Shen 2003; 
Lou \& Zou 2004a, b; Shen \& Lou 2004a, b; Shen, Liu \& Lou 2004; 
Lou \& Wu 2004).

There are various sources of X-ray emissions from galaxies (e.g.,
West, Barber \& Folgheraiter 1997 for the nearby M31 -- the
Andromeda galaxy). X-ray emissions of star-forming galaxies come 
from coronae of stars, close accreting X-ray binaries such as 
low-mass X-ray binaries (LMXBs) and high-mass X-ray binaries (HMXBs), 
supernova remnants and hot plasmas associated with star-forming 
regions as well as galactic winds pumped by starbursts (e.g., 
Fabbiano 1989; Dahlem et al. 1998; Fabbiano \& White 2003; 
Persic \& Rephaeli 2002; Persic et al. 2004; Lou \& Wu 2004).

As some X-ray sources such as HMXBs and Type II supernova remnants
can be physically associated with formation activities of massive 
stars, a correlation may exist between FIR and X-ray global
fluxes/luminosities. In fact, an empirical correlation between 
the FIR and X-ray global luminosities from the young objects (e.g.
HMXBs) has been reported in star-forming galaxies and the correlation
is naturally interpreted in terms of star formation activities in
galaxies, which in turn implies a correlation between X-ray
luminosity and star formation rate (SFR). David et al. (1992)
derived a linear relationship between the logarithms of X-ray
($0.5-4.5$keV) and FIR luminosities, i.e., $L_{\rm 0.5-4.5}\propto
L_{\rm FIR}^{0.95-1.01}$ in the normal and starburst galaxies using 
the X-ray data of {\it EINSTEIN} satellite and suggested a 
two-component model for $L_{\rm 0.5-4.5}$ to fit this correlation. 
Ranalli et al. (2003) inferred a much better correlation between 
soft X-ray ($0.5-2.0$keV)/hard X-ray ($2-10$keV) and FIR/radio 
(1.4GHz) luminosities in star-forming galaxies using X-ray data 
from the {\it BeppoSAX} and {\it ASCA} satellites, and further 
found that such correlation persists in star-forming galaxies of 
higher redshifts $z$ (i.e., $0.2\leq z\leq 1.3$). In particular, 
they suggested that $2-10$ keV X-ray luminosity may be regarded 
as an indicator for the SFR. Grimm, Gilfanov \& Sunyaev (2003) 
used the X-ray luminosity from a collection of HMXBs to estimate 
the SFR and found the correlation between the X-ray luminosity 
and the SFR to be nonlinear in the low SFR regime, while the 
relation becomes linear in the high SFR regime with 
$\hbox{SFR}\geq 4.5 \hbox{M}_{\odot}\hbox{yr}^{-1}$ -- a result 
that differs from that of Ranalli et al. (2003). Gilfanov, Grimm 
\& Sunyaev (2004) discussed the different results and suggested 
that the difference should be caused by X-ray emissions from 
source objects other than HMXBs in the low SFR regime. Persic 
et al. (2004) also used the $2-10$keV X-ray luminosity from a 
collection of HMXBs as an indicator of SFR and pointed out that 
the total $2-10$ keV and soft X-ray luminosities may not be 
reliable SFR indicators.


In this investigation, we shall use the data in soft X-ray
($0.1-2.4$keV) band of {\it R\"ontgen Satellite (ROSAT)} All-Sky
Survey and the FIR ($42.5-122.5\mu$m) data of {\it IRAS} to
examine possible global luminosity correlations in normal
star-forming galaxies and in active star-forming galaxies; the
latter are also classified as starburst-dominant Seyfert galaxies
according to Mouri \& Taniguchi (2002). In particular, the 
correlation between soft X-ray and FIR luminosities in this class 
of galaxies has not been examined before. Also, galaxies in the
data sample we select have not been investigated in the previous
work (e.g., David et al. 1992). We have also acquired the radio
(1.4GHz) data from the NRAO/VLA Sky Survey (NVSS) and the B-band
(B) optical data from both the Lyon/Meudon Extragalactic Database
(LEDA) and the NASA Extragalactic Database (NED). 

\begin{table}
 \centering
 \begin{minipage}{85mm}
  \caption{X-Ray, FIR, Radio (1.4 GHz) and Optical B-Band Data in NSFGs}
  \begin{tabular}{@{}lllllllrlr@{}}
  \hline

   $\it {IRAS}$ Name & $\log{L_{\rm X}}$&$\log{L_{\rm FIR}}$ 
&  $\log{L_{\rm 1.4GHz}}$&$\log{L_{\rm B}}$ & $f_{60}/f_{120}$
     \\
             & \multicolumn{2}{c}{(erg/s)}
&(erg/{s Hz}) & {(erg/s)}&\\
\hline F00146-1934&40.92&43.18&28.17&43.70&0.32\cr
F00317-2142&42.83&44.53&29.82&44.34&0.49\cr F00361-2432
&41.10&43.38 &28.60  &43.58  &0.35\cr F01134+3045 &41.57 &43.36
&29.28 &44.16 &0.41\cr F01590-3158 &41.28  &43.85 &28.92  &43.62
&0.84\cr F02223-1921 &43.02  &44.32  &29.88&44.53  &0.43\cr
F05128+5308 &42.31  &44.38  &29.53 &44.30 &0.74\cr F06483-1955
&42.08  &44.11  &-  &-  &0.71\cr F10290+6517 &40.01  &42.64 &-
&43.43  &0.30\cr F10354-2651 &41.07  &42.85 &- &42.91 &0.45\cr
F15370+3155 &40.67  &42.63 &27.42  &42.84 &0.52\cr F17376+7207
&40.89  &43.20   &-  &43.71 &0.42\cr F17364+2458 &41.48 &43.59  &-
&43.74  &0.64\cr F20240-5233 &41.27  &43.45  &- &43.88 &0.38\cr
F21468-5530 &42.90 &44.13  &- &-  &0.46\cr F23229+2834&42.06
&44.35  &- &44.38 &0.58\cr
\hline
\end{tabular}
\medskip

Column 1 lists names of source objects identified by the {\it IRAS} 
catalogue. Columns 2 through 5 list the logarithms of soft X-ray 
($0.1-2.4$ keV) luminosity, FIR ($42.5-122.5\mu$m) luminosity, radio 
(1.4 GHz) luminosity and optical B-band luminosity, respectively. 
Column 6 lists the FIR colour temperature defined by $f_{60}/f_{120}$.
\end{minipage}
\end{table}

In Section 2, we describe the procedure of data sample selection
and introduce identifications for various types of galaxies. We
analyze available data in Section 3 to reveal correlations between
global soft X-ray and FIR luminosities in normal star-forming
galaxies (NSFGs) (i.e. late-type galaxies and host-dominant
Seyfert galaxies) and in active star-forming galaxies (ASFGs)
(i.e. starburst-dominant Seyfert galaxies); similar analysis fails
to show such a correlation in AGN-dominant Seyfert galaxies.
In parallel, we have also analyzed $L_{\rm X}-L_{\rm B}$,
$L_{\rm FIR}-L_{\rm B}$ and $L_{\rm FIR}/L_{\rm B}-f_{60}/f_{120}$
to examine soft X-ray and star formation properties in NSFGs.
In Section 4, we indicate that multi-wavelength luminosity
correlations may be associated with star formation processes
and the global SFR in galaxies. We infer empirically the relation
between soft X-ray luminosity and the SFR in star-forming galaxies.
In Section 5, we introduce a two-component model to fit the
correlation between soft X-ray and FIR global luminosities 
and estimate relevant errors.

\begin{table}
 \centering
 \begin{minipage}{85mm}
  \caption{X-Ray, FIR, and Optical B-Band Data of Seyfert 1 Galaxies}
  \begin{tabular}{@{}llllllr@{}}
   \hline
     $\it {IRAS}$ Name & $\log{L_{\rm X}}$&$\log{L_{\rm FIR}}$ &
$\log{L_{\rm B}}$ & Notes
        \\
             & \multicolumn{3}{c}{(erg/s)}
& & &\\
\hline F00392-7930     &42.63  &44.18  &44.22   &AGN\cr
F00489+2908     &42.65  &44.15  &44.24    &AGN\cr F01194-0118
&43.39  &44.64  &44.07   &SB\cr F01572+0009     &44.38 &45.75
&44.81  &$m_B>16$\cr F02366-3101     &42.75  &44.53 &44.35 &SB\cr
F02553-1642     &43.56  &44.62  &44.15  &SB\cr F04305+0514 &44.22
&44.23  &44.47   &AGN\cr F04340-1028 &43.68  &44.56 &44.63 &AGN\cr
F04565+0454 &42.09  &43.68 &44.01   &AGN\cr F05136-0012     &44.32
&43.92  &44.59   &AGN\cr F05177-3242 &41.43  &43.37 &43.66 &AGN\cr
F06279+6342 &42.65  &43.47 &43.64  &AGN\cr F06563-6529 &42.89
&43.85 &-  &-\cr F07327+5852     &41.25 &42.58 &42.53  &SB\cr
F07431+6103 &43.49  &43.89  &44.68  &AGN\cr F09497-0122 &41.91
&43.69 &-  &-\cr F09595-0755 &43.63  &44.49 &-  &-\cr F12393+3520
&42.49  &43.91 &44.28  &AGN\cr F12495-1308 &43.01 &43.47 &43.69
 &AGN\cr F13204+0825 &42.62 &44.41 &43.98 &SB\cr
F13224-3809 &43.78 &44.84 &-  &-\cr F15091-2107 &43.02 &44.47
&44.21 &SB\cr F15288+0737 &43.18 &44.05 &44.13  &AGN\cr
F15572+3510 &43.27  &43.89 &44.05  &AGN\cr F17550+6520 &43.51
&44.36 &-  &-\cr F18402-6224 &42.80 &43.06 &43.78   &AGN\cr
F20044-6114 &42.83 &43.47 &44.01 &AGN\cr F20437-0259 &43.30 &43.73
&44.19  &AGN\cr F22062-2803 &43.05 &44.18 &44.45  &AGN\cr
F22402+2927 &44.20 &43.73 &43.97 &AGN\cr F23279-0244 &42.67 &43.96
&44.27  &AGN\cr F08518+1752 &43.58  &44.57 &-  &-\cr 04312+4008
&43.10 &44.07  &-  &-\cr F05262+4432 &42.77 &44.35 &- &-\cr
F09494-0635 &42.47 &43.27  &44.11 &AGN\cr F14400+1539 &44.82
&44.59  &44.77  &$m_B>16$\cr F16136+655 &42.18 &44.83 &43.41
&$m_B>16$\cr F20414-1054 &44.39 &44.19 &44.75  &AGN\cr \hline
\end{tabular}
\medskip

Column 1 lists names of source objects identified by the 
{\it IRAS} catalogue. Columns 2 through 5 list the logarithms 
of X-ray ($0.1-2.4$ keV) luminosity, FIR ($42.5-122.5\mu$m) 
luminosity and optical B-band luminosity, respectively. 
Column 6 lists brief notes: AGN for active galactic nucleus, 
SB for starburst and $m_B$ for brightness magnitude in B-band.

\end{minipage}
\end{table}

\begin{table}
 \centering
 \begin{minipage}{85mm}
  \caption{X-Ray, FIR, and Optical B-Band Data in Seyfert 2 Galaxies}

 \begin{tabular}{@{}lllllll@{}}
   \hline
     $\it {IRAS}$ Name & $\log{L_{\rm X}}$&$\log{L_{\rm FIR}}$ &
$\log{L_{\rm B}}$ & Notes
      \\
             & \multicolumn{3}{c}{(erg/s)}
&
& &\\
\hline F00076-0459&42.94&43.82&44.04&AGN\cr F01413+0205
&41.66&43.57&43.56&SB\cr F02537-1641&42.28&43.92 &43.49&SB\cr
F04265-4801&41.17&43.76&44.02&AGN\cr
F04575-7537&41.66&43.44&43.82&AGN\cr
F05497-0728&41.07&43.42&43.91&AGN\cr F11210-0823
&41.67&43.48&44.08&AGN\cr F13445+1121&41.50&44.21 &43.31
&$m_{B}>16$\cr F14288+5255&42.07&44.29&- &-\cr
F17020+4544&44.10&44.62&-&-\cr F17489+6843 &42.62&44.25&44.14&
SB\cr \hline
\end{tabular}
\medskip

See the same definitions of columns contained in Table 2.
\end{minipage}
\end{table}

\begin{table}
 \centering
 \begin{minipage}{85mm}
  \caption{X-Ray, FIR, and Optical B-Band Data in Seyfert 1.8 Galaxies}
  \begin{tabular}{@{}lllllll@{}}
  \hline
     $\it {IRAS}$ Name & $\log{L_{\rm X}}$&$\log{L_{\rm FIR}}$ &
$\log{L_{\rm B}}$ & Notes
      \\
             & \multicolumn{3}{c}{(erg/s)}
&
& &\\
\hline F14207+3304&41.89&44.04&44.28&host\cr F15083+6825
&42.74&44.49&44.19&SB\cr F22377+0747&43.31&43.76 &43.97& AGN\cr
F22454-1744&44.04&45.07&44.08 &SB\cr \hline
\end{tabular}
\medskip

See the same definitions of columns contained in Table 2. 
Here, `host' stands for a host-dominant Seyfert galaxy.
\end{minipage}
\end{table}

\begin{table}
 \centering
 \begin{minipage}{85mm}
  \caption{X-Ray, FIR, and Optical B-Band Data in Seyfert 1.9 Galaxies }
 \begin{tabular}{@{}lllllll@{}}
   \hline
     $\it {IRAS}$ Name & $\log{L_{\rm X}}$&$\log{L_{\rm FIR}}$ &
$\log{L_{\rm B}}$ & $f_{60}/f_{120}$& Notes
      \\
             & \multicolumn{3}{c}{(erg/s)}
&
& &\\
\hline F11353-4854&42.35&43.76&44.40&AGN\cr F13218-1929
&42.20&43.77&43.38&SB\cr F15564+6359&42.45&43.98 &43.62&SB\cr
F16277+2433&42.05&44.16&43.84&SB\cr F17551+6209 &42.93&44.50&- &
-\cr \hline
\end{tabular}
\medskip

See the same definitions of columns contained in Table 2.

\end{minipage}
\end{table}

\begin{table}
 \centering
 \begin{minipage}{85mm}
  \caption{X-Ray, FIR, and Optical B-Band Data in Seyfert 1.5 Galaxies }
\begin{tabular}{@{}lllllll@{}}
   \hline
     $\it {IRAS}$ Name & $\log{L_{\rm X}}$&$\log{L_{\rm FIR}}$ &
$\log{L_{\rm B}}$ & Notes
      \\
             & \multicolumn{3}{c}{(erg/s)}
&
& &\\
\hline

F01248+1855 &42.99&43.56&43.76&AGN\cr F06457+7429 &42.17
&43.53&43.72&AGN\cr F11034+7250&41.72&43.12 &43.89&AGN\cr
F11112+0951&42.93&44.27&44.44&AGN\cr
F14157+2522&43.68&43.53&44.16&AGN\cr
F23163-0001&43.29&43.97&44.30&AGN\cr

\hline
\end{tabular}

See the same definitions of columns contained in Table 2.
\end{minipage}
\end{table}

\section{SELECTION OF DATA SAMPLES}

Soft X-ray observations of the {\it ROSAT} All-Sky Survey were in
the photon energy band $0.1-2.4$keV. The 120,000 X-ray sources
detected in the RASS II processing of the {\it ROSAT} All-Sky
Survey have been systematically compared and correlated with
14,315 {\it IRAS} galaxies selected from the {\it IRAS} Point
Source Catalogue, among which 197 objects of soft X-ray emissions
are most likely associated with the corresponding {\it IRAS} 
galaxies. For these 197 source objects, their soft X-ray emission 
properties are examined and compared with their FIR emission 
properties. The data in the soft X-ray band are provided by the 
{\it ROSAT} satellite which features a high sensitivity X-ray 
telescope and a low background Position Sensitive Proportional 
Counter (PSPC) detector (see e.g., Boller et al. 1992 and
Boller et al. 1998).

Among these 197 objects, there are 15 spiral galaxies (peculiar
galaxies not included here) and 1 irregular galaxy (see Table 1)
which are classified as late-type galaxies belonging to NSFGs, 38
Seyfert 1 galaxies (see Table 2), 11 Seyfert 2 (see Table 3), 4
Seyfert 1.8 galaxies (see Table 4), 5 Seyfert 1.9 galaxies (see
Table 5) and 6 Seyfert 1.5 galaxies (see Table 6).
The rest of the source objects not tabulated here are elliptical
galaxies, normal galaxies with peculiar characteristics or FIR
sources each corresponding to two or more X-ray sources.

In order to compute a soft X-ray ($0.1-2.4$keV) energy flux from
a PSPC count rate, Boller et al. (1998) presumed a simple power-law
energy spectrum, namely
\begin{equation}
f_{E}\verb"d"E\propto{E^{-\gamma+1}\verb"d"E}\ ,
\end{equation}
where $f_{E}\verb"d"E$ is the energy flux in the photon energy range
of $E$ and $E+\verb"d"E$ from a source galaxy, the fixed photon
spectral index $\gamma$ is taken to be $\sim 2.3$ -- a typical value
estimated for extragalactic source objects found by {\it ROSAT} (see
e.g., Hasinger, Tr\"umper \& Schmidt 1991 and Walter \& Fink 1993),
and the proportional coefficient in expression (1) involves an
absorbing column density of hydrogen fixed at the Galactic value
$N_{\rm{Hgal}}$ along the line of sight (e.g., Stark et al. 1992).

The total FIR ($42.5-122.5\mu$m) flux $f_{\rm FIR}$ may be
estimated (see Helou et al. 1985) by using the $60\mu$ and
$100\mu$ band fluxes, $f_{\rm 60}$ and $f_{\rm 100}$, of
the {\it IRAS} data, namely
\begin{equation}
f_{\rm FIR}\cong 1.26\times 10^{-11}(2.58f_{\rm 60}
+f_{\rm 100})\ \hbox{ erg s}^{-1}\hbox{ cm}^{-2}\ .
\end{equation}
The X-ray and FIR fluxes were converted to luminosities
using the formula adopted by Schmidt \& Green (1986),
namely
\begin{equation}
L(E_{{\rm 1}},E_{{\rm 2}})=4\pi(c/H_{{\rm
0}})^2C(z)A^2(z)f(E_{{\rm 1}},E_{\rm 2})\ ,
\end{equation}
where $c$ is the speed of light,
the Hubble constant $H_{\rm 0}=72$ km s$^{-1}$Mpc$^{-1}$
was recently determined by observations of the Wilkinson 
Microwave Anisotropy Probe ({\it WMAP})
(e.g., Spergel et al. 2003) and the power-law spectrum
in the energy range $(E_{\rm 1}, E_{\rm 2})$ is such
that the two redshift $z$ dependent functions $C(z)$
and $A(z)$ are given by
\begin{equation}
C(z)\equiv (1+z)^{\Gamma-2}\
\end{equation}
and
\begin{equation}
A(z)\equiv 2[(1+z)-(1+z)^{1/2}]\ ,
\end{equation}
respectively. For the photon spectral index $\Gamma$ in the
FIR band, we take $\Gamma=1.5$ (e.g., Boller et al. 1998).
Meanwhile, we have also acquired the radio continuum (1.4 GHz) 
and optical B-band data from the NED and the LEDA.


In this sample, we classify Seyfert galaxies as AGN-dominant
Seyfert galaxies that are characterized by $L_{\rm FIR}/L_{\rm
B}\leq 1$ and $f_{100}/f_{60}\leq 3$, starburst-dominant Seyfert
galaxies that are characterized by $L_{\rm FIR}/L_{\rm B}\geq 1$,
and host-dominant Seyfert galaxies that are characterized by
$f_{100}/f_{60}\geq 3$. We have only selected those galaxies 
with total apparent corrected B-band magnitude $m_B<16$ mag 
(see Mouri \& Taniguchi 2002).


Here we identify 34 AGN-dominant Seyfert galaxies, 14 
starburst-dominant Seyfert galaxies and 1 host-dominant Seyfert 
galaxy. Emissions from AGN-dominant Seyfert galaxies were presumably 
powered by accretion processes around SMBHs in the galactic centers, 
while those from starburst-dominant Seyfert galaxies are powered by 
star formation activities in circumnuclear regions (e.g. Lou et al. 
2001) and are also regarded as ASFGs. For the only host-dominant 
Seyfert galaxy F14207+3304 with the host galaxy being a spiral 
galaxy, we regard it as a NSFG.

On the basis of the preceding classification, we find that
starburst-dominant Seyfert galaxies consist preferentially 
of Seyfert 2 galaxies, while AGN-dominant Seyfert galaxies 
consist preferentially of Seyfert 1 galaxies. From this 
difference, we may be able to test the unification scheme 
of AGN scenario indirectly. In this model scenario, there 
is a dust torus around the central SMBH and the different 
observational appearances of Seyfert 1 and 2 galaxies are 
related to the different geometrical orientations of the 
system relative to the line of sight. Seyfert 1 galaxies 
are thought to be face-on towards us, so that the dust 
torus cannot block or obscure radiations from the central 
SMBH. In this picture, AGN-dominant Seyfert galaxies and 
Seyfert 1 galaxies should be closely related to each other. 
In comparison, for a Seyfert 1 galaxy
nearly edge-on towards us, if the dust torus is Compton 
thick, most of the radiation from the central SMBH will 
be blocked and obscured by the torus with reduced apparent 
AGN emissions. This would make starburst-dominant Seyfert 
galaxies appear as Seyfert 2 galaxies.

\section{DATA ANALYSIS OF GALACTIC LUMINOSITIES}

Here we perform the least-square analysis to reveal possible
correlations in luminosities of galaxies in different bands, 
and provide an estimate for the dispersion or scattering 
around the best-fit relation in terms of $s_1$ for the standard 
deviation $\sigma$, namely
\begin{equation}
s_1\equiv\bigg[\frac{\sum{(\log{L}-\log{L_0})^2}}
{(N-\nu )}\bigg]^{1/2}\ ,
\end{equation}
where $L$ is the observationally inferred luminosity and $L_0$ is 
the luminosity expected from the best fit relation, $N$ is the 
number of sample used and $\nu$ is the number of free parameters. 
In our case, $\nu=2$ and we further introduce a $Q$ parameter for 
the logarithm of the ratio $L_{\rm FIR}$ to $L_{\rm X}$, namely
\begin{equation}
Q\equiv\log{\{[L_{\rm FIR}\hbox{(erg/s)}]/[L_{\rm X}\hbox{(erg/s)}]\}}\ .
\end{equation}
This $Q$ parameter is analogous to the suggested $q$ parameter by
Helou et al. (1986) for the correlation between $L_{\rm FIR}$ and
$L_{\rm RADIO}$ and we here analyze the correlation between soft
X-ray and FIR luminosities, and the standard deviation of $Q$ can
be similarly defined by
\begin{equation}
s_2\equiv\bigg[\frac{\sum{(Q_i-\bar{Q})^2}}{(n-1)}\bigg]^{1/2}\ ,
\end{equation}
where $\bar{Q}$ is the mean $Q$ and $n$ is the number of data sample.

\begin{figure}
\begin{center}
\includegraphics[scale=0.45]{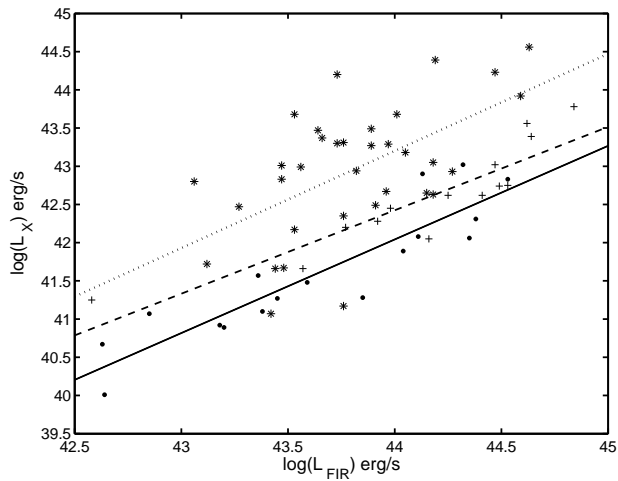}
\caption{Soft X-ray ($0.1-2.4$keV) luminosity $L_{\rm X}$ versus
FIR ($42.5-122.5\mu$m) luminosity $L_{\rm FIR}$ in a log-log plot
for star-forming galaxies and AGN-dominant Seyfert galaxies
together. Solid circles represent data points of NSFGs, crosses
represents data points of ASFGs and asterisks represent data 
point of AGN-dominant Seyfert galaxies. The straight solid line
represents a least-square linear fit for NSFGs, the straight
dashed line represents a least-square linear fit for ASFGs and 
the straight dotted line represents a least-square linear fit 
for AGN-dominant Seyfert galaxies).
This plot is intended to compare the correlations in different
classes of galaxies and to test the suggestion that some LLAGNs
might be more controlled by the star-forming process. }
\end{center}
\end{figure}

\begin{figure}
\begin{center}
\includegraphics[scale=0.45]{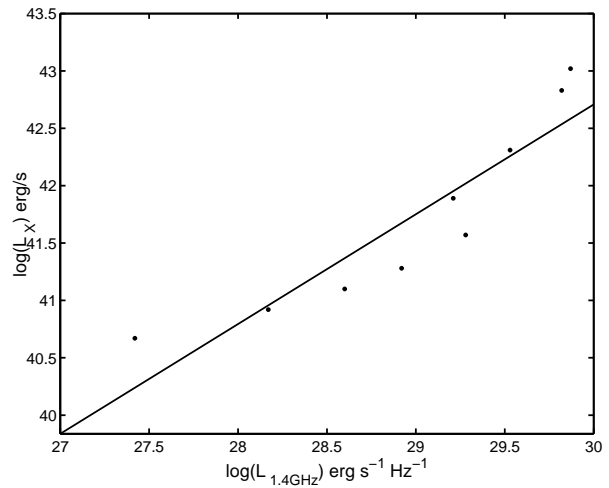}
\caption{Soft X-ray ($0.1-2.4$keV) luminosity $L_{\rm X}$ versus
radio (1.4 GHz) luminosity $L_{\rm 1.4GHz}$ in a log-log plot for
NSFGs. Solid circles are data points for NSFGs and the straight
solid line represents a least-square linear fit for NSFGs. }
\end{center}
\end{figure}


\begin{table*}
 \centering
 \begin{minipage}{128mm}
  \caption{The main results of multi-wavelength luminosity 
correlations in NSFGs, ASFGs, AGN-dominant Seyfert (AGN 
for short in this table) galaxies}
\begin{tabular}{@{}llrllllll@{}}
   \hline
     $Y$&\ $X$&GS\ \ \ \ &\ \ \ $m$&\ \ \ \ $c$&\ \ $r$&
\ \ $s_{1}$&\ \ $\bar{Q}$&$\ \ s_{2}$
      \\

\hline

$\log L_{\rm X}$&$\log L_{\rm FIR}
$&NSFGs&$1.22\pm0.15$&$-11.85\pm6.48$&0.91&0.37&2.04&0.38\cr
$\log L_{\rm X}$&$\log L_{\rm1.4GHz}$
&NSFGs&$0.95\pm0.14$&$-13.94\pm6.01$&0.92&0.36&-&-\cr
$\log L_{\rm X}$&$\log L_{\rm FIR}
$&ASFGs&$1.09\pm0.15$&$-5.63\pm6.65$&0.90&0.32&1.56&0.38\cr
$\log L_{\rm X}$&$\log L_{\rm FIR}
$&AGNs&$1.27\pm0.07$&$-12.74\pm2.79$&0.57&0.71&0.82&0.71\cr
$\log L_{\rm X}$&$\log L_{\rm B}
$&NSFGs&$1.26\pm0.25$&$-14.08\pm11.34$&0.80&0.50&-&-\cr
$\log L_{\rm FIR}$&$\log L_{\rm B}$
&NSFGs&$1.08\pm0.17$&$-3.56\pm7.38$&0.87&0.33&-&-\cr

\hline
\end{tabular}

Here we assume a linear fitting expression in the form of 
$Y=mX+c$ and we summarize the $Y$, $X$ physical variables, 
galaxy samples (GS), the fitting slope $m$, the fitting 
intercept $c$, the correlation coefficient $r$, the 
standard deviation of the correlation $s_{1}$, the mean 
of $Q$ parameter $\bar{Q}$, and the standard deviation 
$s_{2}$ in $Q$ parameter. 
\end{minipage}
\end{table*}

\subsection{Correlations of Global FIR (Radio)
and X-Ray Luminosities from Galaxies}

Displayed in Figure 1 are soft X-ray ($0.1-2.4$keV) luminosity
$L_{\rm X}$ versus FIR luminosity $L_{\rm FIR}$ of NSFGs (solid
circles), ASFG (crosses) and AGN-dominant galaxies (asterisks),
respectively, in a log-log plot to confirm the correlations between 
soft X-ray and FIR global luminosities, and their least-square fit 
lines are straight solid, dashed and dotted lines, respectively. 
We list the results of $L_{\rm X}-L_{\rm FIR}$ correlation and the 
mean value of $Q$ in Table 7. The $Q$ value for the source object 
F14207+3304 is $\simeq 2.15$ suggesting that F14207+3304 belongs 
to the class of NSFGs.

We notice the difference of the correlation in NSFGs and ASFGs, 
that is, the mean value of $\bar Q$ in NSFGs is larger than that 
in ASFGs. For the same soft X-ray luminosity, the global FIR
emission from the ASFGs is weaker than that from the NSFGs. Or
equivalently, for the same global FIR luminosity, the X-ray
luminosity from the NSFGs is weaker than that from the ASFGs.

In reference to the luminosity correlations in star-forming 
galaxies (including NSFGs and ASFGs), such a correlation 
in AGN-dominant Seyfert galaxies becomes much weaker with 
considerable scatters.
By using the F-test (e.g., Taylor 1981), we find that
the correlations in the NSFGs and the ASFGs are much better than
that in the AGN-dominant Seyfert galaxies at $\alpha=0.01$ level.
And the mean value of $Q$ in the AGN-dominant Seyfert galaxies is 
much lower than those in the NSFGs and the ASFGs; this is most likely
caused by central accretion activities around SMBHs harboured in 
galactic nuclei. By a visual inspection of Fig. 1, it is clear that 
for the same FIR luminosity, the soft X-ray emission is considerably
higher in AGN-dominant Seyfert galaxies. Also from Fig. 1, we note 
that the distribution of AGN-dominant Seyfert galaxies is bound 
from below by a relatively narrow region occupied by star-forming
galaxies. Some LLAGNs also show a similar correlation that we infer 
for star-forming galaxies, implying that emission processes from 
these LLAGNs might be more controlled by star formation activities.

In addition, we infer a correlation between soft X-ray and 1.4GHz 
radio continuum luminosities in NSFGs as shown in Fig. 2; details 
of this empirical relation are summarized in Table 7.
\begin{figure}
\begin{center}
\includegraphics[scale=0.45]{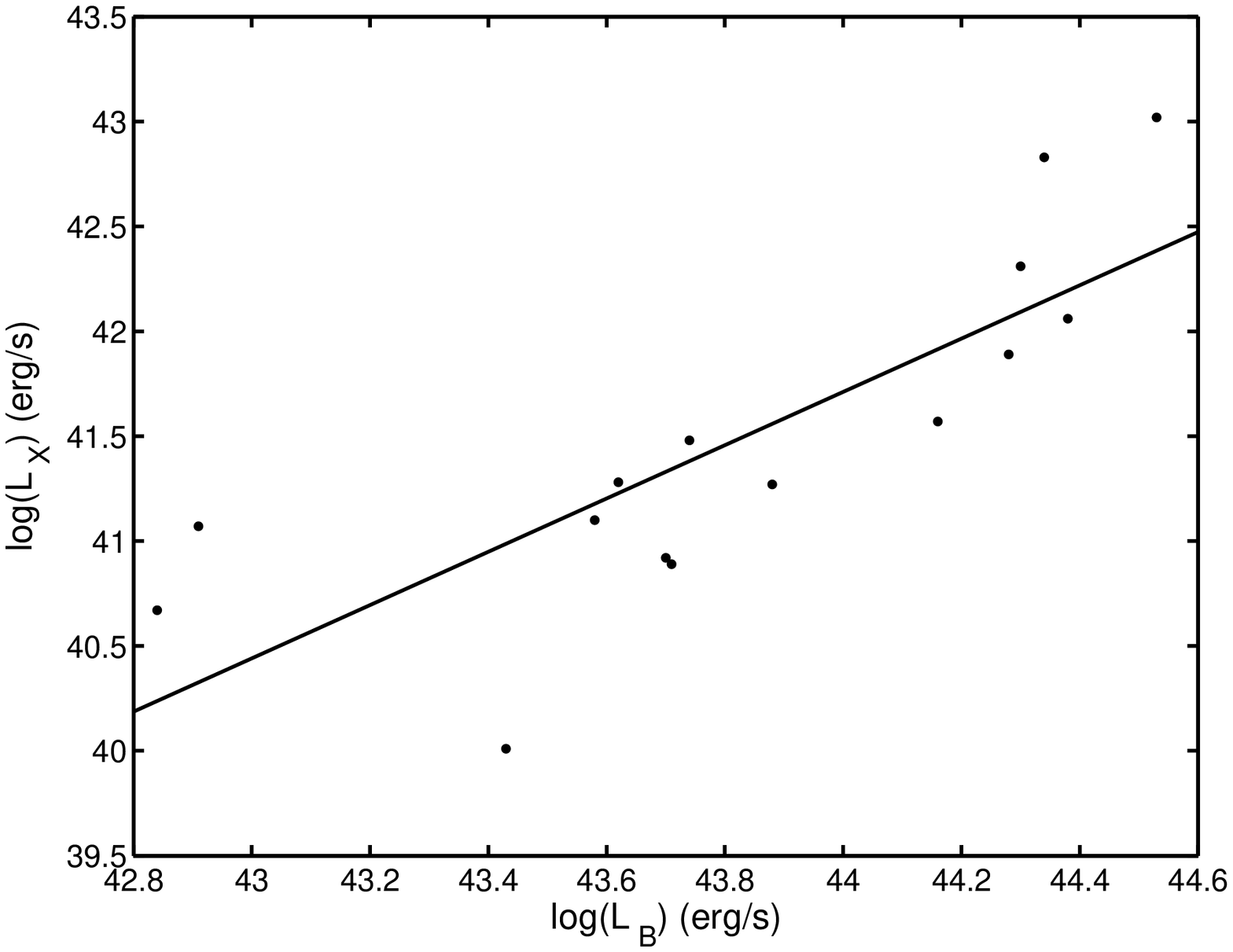}
\caption{Soft X-ray ($0.1-2.4$keV) luminosity $L_{\rm X}$ versus
optical B-band luminosity $L_{\rm B}$ of NSFGs in a log-log plot.
Solid circles represent data points of NSFGs. The straight solid
line is a least-square linear fit for NSFGs.  }
\end{center}
\end{figure}

\begin{figure}
\begin{center}
\includegraphics[scale=0.45]{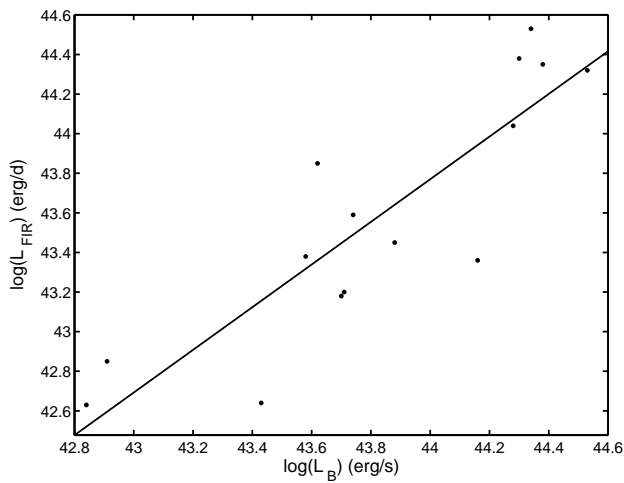}
\caption{FIR ($42.5-122.5\mu$m) luminosity $L_{\rm FIR}$ versus
optical B-band luminosity $L_{\rm B}$ of NSFGs in a log-log plot.
Solid circles represent data points of NSFGs. The straight solid
line is a least-square linear fit for NSFGs. }
\end{center}
\end{figure}

\begin{figure}
\begin{center}
\includegraphics[scale=0.45]{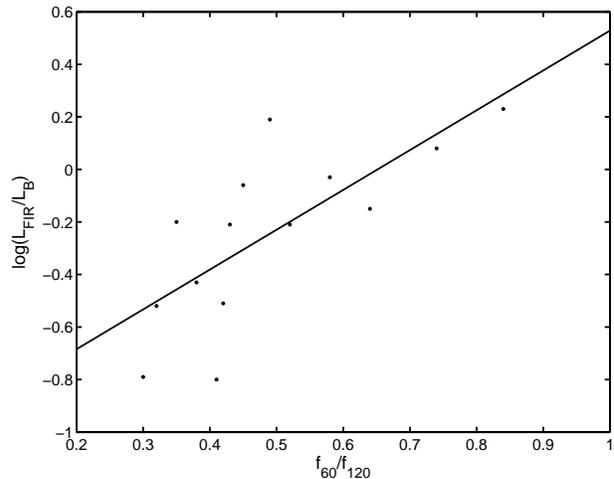}
\caption{Colour temperature $f_{60}/f_{120}$ of late-type 
galaxies versus $\log(L_{\rm FIR}/L_{\rm B}$). Solid dots 
represent data points of late-type galaxies. The straight 
solid line is a least-square linear fit. }
\end{center}
\end{figure}

\subsection{Correlations of X-Ray (FIR) and Optical 
B-Band Luminosities in Star-Forming Galaxies}

In the previous section, we have inferred the correlations between
soft X-ray and FIR global luminosities in star-forming galaxies.
In Figs. 3 and 4 of this section, we examine the $L_{\rm X}$ 
versus $L_{\rm B}$ and $L_{\rm FIR}$ versus $L_{\rm B}$ relations 
in the NSFGs, with details of this relation contained in Table 7.

Except at the $\alpha=0.25$ level, the correlation in soft 
X-ray and FIR luminosities is not significantly better 
than that in soft X-ray and optical B-band luminosities. 
In star-forming galaxies, the more massive the galaxy is, the 
more regions of star formation it will contain, because the 
optical B-band luminosity may roughly represent the mass of a 
galaxy. The X-ray/FIR luminosities can be determined by star 
formation activities and this may be the underlying cause for 
the correlation between X-ray/FIR and optical B-band emissions. 
It is sensible to reason that the star formation rate is not 
only associated with the mass of a galaxy but also with the 
active level of star formation processes in that galaxy. The
active level of star formation processes in galaxies can be
roughly represented by both $L_{\rm FIR}/L_{B}$ and the FIR 
colour temperature $f_{60}/f_{120}$; their gross correlation 
is therefore shown in Fig. 5 with a correlation coefficient 
$r\simeq 0.73$.

\subsection{Evolution from Starburst Galaxies
to Starburst-Dominant Seyfert Galaxies}

In Fig. 1, we call attention to a quantitative difference 
in global luminosity correlations in NSFGs and ASFGs. This 
difference can also be expressed in terms of different mean 
values of $\bar Q$, that is, for the same global soft X-ray 
luminosity, the FIR emission tends to be stronger in NSFGs, 
or equivalently, for the same global FIR luminosity, the soft 
X-ray emission tends to be stronger in ASFGs. We suspect 
that this difference might be related to an evolution from 
starburst galaxies to starburst-dominant Seyfert galaxies.

The physical process of starbursts is thought to be a key 
piece of the AGN machinery. The central AGN and the process 
of starbursts may be either symbiotic or evolutionary (e.g., 
Perry \& Dyson 1985).
Circumnuclear starbursts occur at radii of $\sim 10^2-10^3$ pc,
while the fueling of an AGN occurs at radii of $\ll 1$ pc. More 
observational evidence and theoretical studies (e.g., Lou et al. 
2001 and references therein) point to the possibility of MHD density 
waves as the underlying cause of large-scale ``ring" or spiral as 
well as other patterns. In some cases, these circumnuclear patterns 
are likely driven by galactic bars on even larger scales (e.g.,
NGC 1097). From the evolutionary point of view, it is conceivable 
that a young starburst galaxy may gradually evolve towards a 
maximum phase of starburst-dominant Seyfert galaxy and then decay 
as starburst activities gradually peter out.


The difference in correlations between $L_{\rm X}$ and 
$L_{\rm FIR}$ in the NSFGs and in the ASFGs as observed may reflect 
the two phases along one evolutionary track. Circumnuclear starbursts 
in Seyfert galaxies are thought to be sustained over a relatively
long period of time. Their starburst ages might be on the order of
$\sim 10^8$yr -- approximately 10 times longer than the typical
classical starburst age (e.g., Glass \& Moorwood 1985). In cases
of relatively long-lived starburst activities, stellar winds and
supernova explosions might have driven away considerable amount of
dust grains that would be responsible for FIR emissions, and UV 
photons from young massive stars may then travel longer distances 
before being absorbed by dust grains (e.g., Mouri \& Taniguchi 
2002). In this scenario, the FIR luminosity will decrease 
leading to a difference in the correlation of $L_{\rm X}$ 
and $L_{\rm FIR}$ and thus lower values of $Q$.

\section{SOFT X-RAY LUMINOSITY AND STAR FORMATION RATE (SFR)}

The SFR is one of the key parameters to characterize the formation
and evolution of galaxies. Some empirical relations have been
found between global luminosities in different wave bands and SFR
in galaxies (see some examples in Table 3 of Grimm et al. 2003).
In particular, Kennicutt (1998) inferred a correlation between 
the FIR and SFR through a set of numerical simulations, namely
\begin{equation}
\hbox{SFR}=\frac{L_{\it FIR}}{2.2\times10^{43}}
\hbox{M}_{\odot}/\hbox{yr}\ .
\end{equation}
Condon (1992) derived an empirical relation 
between the radio continuum luminosity at 1.4 GHz 
and the SFR in galaxies in the form of
\begin{equation}
\hbox{SFR}=\frac{L_{\rm 1.4GHz}}{4.0\times10^{28}}
\hbox{M}_{\odot}/\hbox{yr}\ .
\end{equation}
The existence of empirical correlations in the logarithms of
FIR/radio and soft X-ray global luminosities is naturally
suggestive that these emission bands may contain physically
relevant information of star formation activities. There are a 
few more plausible reasons for doing so. In terms of statistics, 
galactic X-ray emissions are more abundant in soft bands. While 
soft X-ray emissions may suffer random scatters, they are less 
affected by contributions of non-star-formation origins, including 
for example Compton-thick AGNs. We therefore tentatively propose a 
relation between the SFR and the soft X-ray ($0.1-2.4$keV) 
luminosity in reference to $L_{\rm FIR}$. On the basis of a mean 
$\bar Q\simeq 2.04$ for NSFGs and equation (9), we then infer
\begin{equation}
\hbox{SFR}=\frac{L_{\rm X(0.1-2.4keV)}}
{2.0\times10^{41}}\hbox{M}_{\odot}/\hbox{yr}\ .
\end{equation}
Using the similar procedure, we can derive another SFR$-L_{\rm X}$
relation based on correlation (10) between soft X-ray and radio
global luminosities,
\begin{equation}
\hbox{SFR}=\frac{L_{\rm X(0.1-2.4keV)}}{2.2\times10^{41}}
\hbox{M}_{\odot}/\hbox{yr}\ ,
\end{equation}
which is fully compatible with equation (9) as expected. By
equations (9) or (10), one may empirically estimate SFRs via 
soft X-ray ($0.1-2.4$keV) global luminosities in star-forming 
galaxies. In other words, the soft X-ray luminosity in the 
energy band $0.1-2.4$keV may serve as a SFR indicator.

Whether the global soft X-ray luminosity is a sure SFR indicator in 
a galaxy still remains an interesting open question. In a recent 
work of data analysis (e.g. Persic et al. 2004), as scatters in the 
$L_{\rm X}-$SFR relation is somewhat higher in the soft band than 
in the hard band, the soft X-ray global luminosity was regarded as 
a poor indicator for the SFR. Persic et al. (2004) noted that part 
of these scatters was due to observational errors and most scatters 
might be caused by X-ray emissions from SN-powered outgoing galactic 
winds.\footnote{For a recently suggested concept of a {\it spiral 
galactic wind} away from both sides of a galactic disc plane, the reader 
is referred to the global construction of stationary MHD perturbation 
patterns in a composite system of a stellar disc and an isopedically 
magnetized gaseous disc by Lou \& Wu (2004). Open magnetic field 
lines and disc activities facilitate ecapes of hot gases. Large-scale 
MHD density waves lead to large-scale spiral galactic winds.} 
Although a galactic wind may also be associated with star formation 
processes (e.g., even as an important signature of starbursts;
Dahlem et al. 1998; Persic \& Rephaeli 2002; Persic et al. 2004), 
the luminosity of the wind is also influenced by various local 
source properties. Tyler et al. (2004) found that there exists a 
fairly good correlation between the level of diffuse soft X-ray 
emissions and the level of mid-infrared emissions in different 
regions along spiral arms; such a correlation between diffuse soft 
X-ray and mid-IR fluxes around galactic centers becomes less clear, 
tending to support the conclusion of Persic et al. (2004).
On the other hand, Ranalli et al. (2003) get less scatters
in $L_{\rm X}-L_{\rm FIR}$ correlation in soft X-ray band.
This does not rule out the possibility that intrinsic absorptions
may cause considerable scatters in some cases. In short, global 
soft X-ray luminosity may seem to serve as an indicator for the SFR 
in star-forming galaxies in some cases (e.g., Compton thick AGNs or 
the diffuse X-ray luminosity in the center caused by SN explosions 
is much weaker than that from the spiral arms.)

\begin{figure}
\begin{center}
\includegraphics[scale=0.50]{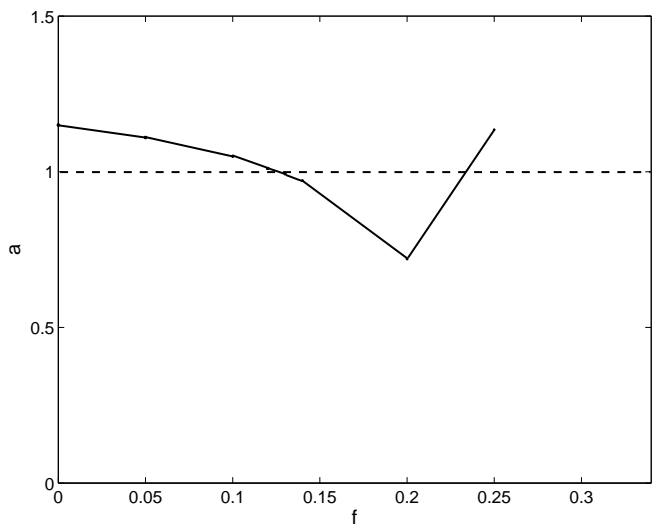}
\caption{The fitting slope $a$ varies with the fitting parameter
$f$ ($f=0, 0.05, 0.15, 0.2, 0.25$). We pick the value of $f$ for
$a=1$. There are two values of $f$ satisfying this criterion. We
choose the smaller value of $f$ as the correlation coefficient is
higher.
}
\end{center}
\end{figure}

\section{TWO-COMPONENT MODEL}

For NSFGs, we propose an empirical model that both FIR and soft
X-ray global luminosities consist of two components, respectively 
(e.g., David et al. 1992). For FIR emissions, one component is 
warmer and is presumably originated from active regions of star 
formation processes, and the other component is cooler and might 
be cirrus-like dust conglomerations heated by an interstellar 
radiation field within a galactic disc that are only weakly 
associated with star formation processes. For soft X-ray emissions, 
one component of radiation comes from relatively young objects, 
including HMXBs, Type II supernovae, hot plasmas associated with
regions of active star formation and with hot galactic winds from
circumnuclear starburst rings as well as possibly from luminous
spiral arms (Lou \& Wu 2004). These sources usually involve 
massive stars that heat dusts and grains to radiate FIR bands 
during certain evolution stages. This may be the physical origin 
for the correlation between FIR and soft X-ray global luminosities. 
The other component of soft X-ray radiation comes from relatively 
older objects, including LMXBs and Type I supernovae; a correlation 
between this component and star-formation activities is thought 
to be relatively weak.

Based on these physical ideas, we propose to decompose 
$L_{\rm FIR}$ and $L_{\rm X}$ in simple forms of
\begin{equation}
L_{\rm FIR}=L_{\rm FIRsf}+L_{\rm FIRnsf}
\end{equation}
and
\begin{equation}
L_{\rm X}=L_{\rm Xsf}+L_{\rm Xnsf}\ ,
\end{equation}
where $L_{\rm FIRsf}$ and $L_{\rm Xsf}$ are respectively FIR and
soft X-ray luminosities that are more closely associated with 
star formation processes (hence the added subscript `sf'), 
$L_{\rm FIRnsf}$ and $L_{\rm Xnsf}$ are respectively FIR and X-ray
luminosities that belong to the component without associations to
star formation processes (hence the added subscript `nsf'). The
life time of source objects without association to star formation
(e.g., LMXBs etc.) might be comparable to the life time of the host
galaxy. Consequently, their luminosities may be proportional to
the total stellar mass of the respective host galaxies which can
be represented by the B-band luminosity (e.g., Grimm et al. 2002;
Gilfanov et al. 2004). In reference to optical B-band luminosities
of galaxies,
we presume $L_{\rm FIRnsf}=fL_B$ and $L_{\rm Xnsf}=gL_B$ where $f$
and $g$ are two numerical fitting factors to characterize FIR and
soft X-ray luminosities of non-star-formation origins in
respective proportions to the B-band luminosity.
In the NSFGs, global radio continuum emissions at 1.4 GHz frequency 
are thought to be primarily caused by star formation processes (e.g., 
relativistic cosmic-ray electrons gyrating in magnetic field). Based 
on earlier empirical results, we aim at fitting $L_{1.4GHz}$ and 
$L_{\rm FIRsf}$ (i.e., $L_{\rm FIR}-fL_B$) with a linear correlation. 
Taking $L_{\rm FIRsf}$ as an example, our procedure of fitting is as 
follows. We write
$$
\log L_{1.4GHz}=a\log (L_{\rm FIR}-fL_B)+b\ ,
$$
where $a$ is the slope and $b$ is the intersection in a log-log
plot. For a trial value of $f$, we input data for $L_{1.4GHz}$,
$L_{\rm FIR}$ and $L_B$ to search for a least-square fit
parameters $a$ and $b$. We then plot the variation of $a$ (i.e.,
the least-square slope $a$) for a set of $f$ values as shown in
Fig. 6.
For $a=1$, we have the $L_{1.4GHz}$ proportional to $L_{\rm
FIRsf}$ linearly. In reference to Fig. 9, we may estimate the
value of $f\simeq 0.13$ that is compatible with the result
$f\simeq 0.14$ derived earlier by Devereux \& Eales (1989).
%

Using the same method of estimation, we have inferred a value of
$g$ to be $\sim10^{-3}$ for $L_{\rm Xsf}\propto L_{\rm FIRsf}$.

On the basis of expressions (13), (14) and 
$L_{\rm Xsf}=d L_{\rm FIRsf}$, $L_{\rm X}$ 
may be expressed by the linear relation of
$L_{\rm FIR}$ and $L_{\rm B}$,
\begin{equation}
L_{\rm X}=AL_{\rm FIR}+BL_{\rm B}\ ,
\end{equation}
where $A=d$ and $B=g-df$. Using the sample of NSFGs and 
performing a $\chi^2$ minimization, we derive the best fit as
\begin{equation}
L_{\rm X}=6.4\times10^{-3}L_{\rm FIR}+3.84\times10^{-4}L_{\rm B}
\end{equation}
for equation (8), where we infer again a value of $g\sim 10^{-3}$.

\section{SUMMARY AND PERSPECTIVE }

We have analyzed a sample of 17 NSFGs, 14 
ASFGs and 34 AGN-dominant Seyfert galaxies,
taken from {\it ROSAT} RASS II observations of {\it IRAS} 
galaxies. We infer different $L_{\rm X}-L_{\rm FIR}$ 
correlations in these groups of galaxies and discuss 
plausible causes for these differences. The difference 
between NSFGs and ASFGs might be caused by an evolution from 
classic starburst to starburst-dominant Seyfert galaxies, 
while the difference between NSFGs/ASFGs and AGN-dominant 
Seyfert galaxies is likely to be caused by the excess of 
X-ray emissions from the nuclei of galaxies. And the 
distribution of the AGN-dominant Seyfert galaxies implies 
that some LLAGNs might be more dominantly influenced by 
star formation processes.

In addition, we also estimate 
the $L_{\rm X}-L_{\rm 1.4GHz}$, 
$L_{\rm X}-L_{\rm B}$, $L_{\rm FIR}-L_{\rm B}$, 
$L_{FIR}/L_{B}-f_{60}/f_{120}$ relations in NSFGs to 
complement our study of the correlation between the FIR 
and soft X-ray global luminosities as well as the relation 
between such correlation and star formation activities.

Condon (1992) and later Kennicutt (1998) inferred the 
SFR$-L_{\rm 1.4GHz}$ and SFR$-L_{\rm FIR}$ relations,
respectively. The correlations among FIR, radio continuum 
and soft X-ray luminosities studied here then imply that 
FIR, radio continuum and soft X-ray luminosities come from 
physical processes involved in star formation activities. 
We thus propose an empirical relation between the SFR and the 
soft X-ray global luminosity in star-forming galaxies. As an 
additional diagnostics, it is then possible to estimate SFRs 
in star-forming galaxies from soft X-ray global luminosities.

In order to plausibly account for these empirical luminosity
correlations and deviations for galaxies, we advance a
two-component model in this paper: one component of FIR (soft 
X-ray) emission is associated with star formation processes, 
while the other component is not.

On the basis of this model analysis, we infer a linear correlation
between FIR/X-ray and radio global luminosities from star
formation regions, and express soft X-ray luminosity $L_{\rm X}$
as a linear combination of $L_{\rm FIR}$ and $L_{\rm B}$. By two
different yet complementary ways, we infer almost the same value
of $g$ factor.

In our data samples, the redshift $z$ of selected galaxies are
fairly small with majority of them being $z\sim 0.05$. Recently,
two of the NASA's Great Space Observatories [i.e., Chandra X-ray
Observatory ({\it CXO}) and Hubble Space Telescope ({\it HST})]
bolstered by the largest ground-based telescopes including the
Very Large Telescope ({\it VLT}), Keck Telescope, Gemini Telescope
and National Optical Astronomy Observatory ({\it NOAO}) around the
world, are beginning to harvest new clues to the origin and
evolution of galaxies in the universe. Barger et al. (2003)
obtained optical and near infrared properties for the 2 ms {\it
Chandra} deep field north X-Ray Sources using the data of {\it Subaru}
--- an 8.2m telescope operated by the National Astronomical
Observatory Japan ({\it NAOJ}). In particular, the Spitzer Space
Telescope (previously known as {\it SIRTF}) launched last year has
already joined the survey referred to as the Great Observatories
Origins Deep Survey ({\it GOODS}). By higher sensitivity, better
pointing accuracy and broad-band spectral capabilities of the {\it
Spitzer}, {\it CXO} and {\it HST}, the {\it GOODS} survey will
provide unprecendented opportunities for more systematic
investigations on interrelations and classifications among
optical, FIR and X-ray global luminosities of galaxies in the 
deep field of higher redshifts ($z\sim 1-2$), where earlier star
formation can be probed. The empirical correlation between
far-infrared and soft X-ray luminosities/fluxes inferred here for
fairly low $z$ will then be further tested for higher $z$.

\section*{Acknowledgments}

We thank the anonymous referee for valuable constructive
suggestions to improve the manuscript. This research was supported
in part by the ASCI Center for Astrophysical Thermonuclear Flashes
at the University of Chicago under Department of Energy contract
B341495, by the Special Funds for Major State Basic Science
Research Projects of China, by the Tsinghua Center for
Astrophysics, by the Collaborative Research Fund from the National
Natural Science Foundation of China (NSFC) for Young Outstanding
Overseas Chinese Scholars (NSFC 10028306) at the National
Astronomical Observatory, Chinese Academy of Sciences, by NSFC
grant 10373009 at the Tsinghua University, and by the Yangtze
Endowment from the Ministry of Education through the Tsinghua
University. Affiliated institutions of Y.Q.L. share the
contribution.

\bsp

\label{lastpage}
\end{document}